\RequirePackage{ifpdf}
\ifpdf 
\documentclass[pdftex]{sigma}
\else
\documentclass{sigma}
\fi

\begin{document}

\allowdisplaybreaks

\renewcommand{\thefootnote}{$\star$}

\renewcommand{\PaperNumber}{031}

\FirstPageHeading

\ShortArticleName{Dif\/ferential and Functional Identities for the Elliptic Trilogarithm}

\ArticleName{Dif\/ferential and Functional Identities \\ for the Elliptic Trilogarithm\footnote{This paper is a contribution to the Proceedings of the Workshop ``Elliptic Integrable Systems, Isomonodromy Problems, and Hypergeometric Functions'' (July 21--25, 2008, MPIM, Bonn, Germany). The full collection
is available at
\href{http://www.emis.de/journals/SIGMA/Elliptic-Integrable-Systems.html}{http://www.emis.de/journals/SIGMA/Elliptic-Integrable-Systems.html}}}

\Author{Ian A.B. STRACHAN}

\AuthorNameForHeading{I.A.B.~Strachan}

\Address{Department of Mathematics, University of Glasgow, Glasgow G12 8QQ, UK}
\Email{\href{mailto:i.strachan@maths.gla.ac.uk}{i.strachan@maths.gla.ac.uk}}
\URLaddress{\url{http://www.maths.gla.ac.uk/~iabs/}}

\ArticleDates{Received November 25, 2008, in f\/inal form March 06,
2009; Published online March 13, 2009}

\Abstract{When written in terms of $\vartheta$-functions, the classical Frobenius--Stickelberger pseudo-addition formula
takes a very simple form. Generalizations of this functional identity are studied, where the functions
involved are derivatives (including derivatives with respect to the modular parameter) of the
elliptic trilogarithm function introduced by Beilinson and Levin. A dif\/ferential identity satisf\/ied
by this function is also derived.
These generalized Frobenius--Stickelberger identities play a fundamental role in the development of elliptic solutions of the Witten--Dijkgraaf--Verlinde--Verlinde
equations of associativity, with the simplest case reducing to the above mentioned dif\/ferential identity.}


\Classification{11F55; 53B50; 53D45}

\renewcommand{\thefootnote}{\arabic{footnote}}
\setcounter{footnote}{0}

\section{Introduction}

Amongst the many beautiful identities satisf\/ied by the elliptic (and related)
functions is the Frobenius--Stickelberger relation \cite{FS,W}
\begin{gather}
\left(\zeta(a)+\zeta(b)+\zeta(c)\right)^2 = \wp(a)+\wp(b)+\wp(c),\qquad (a+b+c=0).
\label{fs}
\end{gather}
Writing this identity not in terms of Weierstrass functions but in terms of
$\vartheta$-functions yields the equivalent form
\begin{gather}\left[
\frac{\vartheta_1'(a)}{\vartheta_1(a)}
\frac{\vartheta_1'(b)}{\vartheta_1(b)}+
\frac{\vartheta_1'(b)}{\vartheta_1(b)}
\frac{\vartheta_1'(c)}{\vartheta_1(c)}+
\frac{\vartheta_1'(c)}{\vartheta_1(c)}
\frac{\vartheta_1'(a)}{\vartheta_1(a)}\right]+ \frac{1}{2} \left[
\frac{\vartheta_1''(a)}{\vartheta_1(a)}+
\frac{\vartheta_1''(b)}{\vartheta_1(b)}+
\frac{\vartheta_1''(c)}{\vartheta_1(c)}\right]=
\frac{1}{2} \frac{\vartheta_1'''(0)}{\vartheta_1'(0)} ,
\label{thetaidentity}
\end{gather}
where again $a+b+c=0$.
With the help of the heat equation the second set of terms may be written in terms of
derivatives with respect to the modular parameter $\tau$.

The purpose of this paper is to explore certain neo-classical identities
satisf\/ied by the elliptic trilogarithm introduced by Beilinson and Levin \cite{BL,Levin}. This function
will be def\/ined in Section~\ref{section2}, but for now it is suf\/f\/icient to note that
with it the above identity takes the simplif\/ied form
\[
\left\{
\begin{array}{c}
\phantom{+}f^{(3,0)}(a) f^{(3,0)}(b)\vspace{1mm}\\
+f^{(3,0)}(b) f^{(3,0)}(c)\vspace{1mm}\\
+f^{(3,0)}(c) f^{(3,0)}(a)
\end{array}
\right\}
-\big\{ f^{(2,1)}(a) + f^{(2,1)}(b) + f^{(2,1)}(c)\big\}=0 .
\]
These new identities take the schematic form
\begin{gather}
\left\{\text{quadratic~terms}\right\} + \left\{\text{linear~terms} \right\} = \left\{ \vartheta~\text{constants} \right\},
\label{genform}
\end{gather}
where the linear terms contain one more $\tau$-derivative than the total number of $\tau$-derivatives
in each part in the quadratic term. Before these identities are discussed (Section~\ref{section4}) a dif\/ferential
equation satisf\/ied by this function $f(z,\tau)$ will be derived in Section~\ref{section3}. Applications of these
new identities are then discussed in Section~\ref{section5}. We begin by def\/ining the elliptic polylogarithm.

\section{The elliptic polylogarithms}\label{section2}

The classical polylogarithm is def\/ined, for $|z|<1$, by
\[
{\rm Li}_r(z) = \sum_{n=1}^\infty \frac{z^n}{n^r}
\]
and by analytic continuation elsewhere. A f\/irst attempt at an elliptic
analogue of this function might be
\[
{\rm \mathcal{L}i}_r(\zeta,q) = \sum_{n=-\infty}^\infty {\rm Li}_r(q^n \zeta).
\]
However this series diverges, but by using the inversion formula
(\ref{polyloginversion}) and $\zeta$-function regularization one can
arrive at the following def\/inition of the elliptic polylogarithm
function \cite{BL,Levin}:
\[
{\rm \mathcal{L}i}_r(\zeta,q) = \sum_{n=0}^\infty {\rm Li}_r(q^n \zeta) +
\sum_{n=1}^\infty {\rm Li}_r\big(q^n \zeta^{-1}\big) -
\chi_r(\zeta,q),\qquad  r~{\rm odd} ,
\]
where
\[
\chi_r(\zeta,q)=\sum_{j=0}^r \frac{B_{j+1}}{(r-j)!(j+1)!}
(\log\zeta)^{(r-j)} (\log q)^j .
\]
A real-valued version of this function had previously been studied by Zagier \cite{Z}.
With this the function $f$ may be def\/ined.

\begin{definition}
\label{defnf}
The function $f(z,\tau),$ where $z\in\mathbb{C}$, $\tau\in\mathbb{H}$, is def\/ined
to be:
\[
f(z,\tau)=\frac{1}{(2 \pi i)^3} \big\{{\rm \mathcal{L}i}_3\big(e^{2 \pi i z},q\big)-{\rm \mathcal{L}i}_3(1,q)\big\} .
\]
\end{definition}

 The function $f^{(n,m)}$ that appear in the introduction
are the derivatives
\[
f^{(n,m)} = \frac{\partial^{n+m} f}{\partial z^n \partial \tau^m} .
\]
It immediately follows from the def\/inition that
\begin{gather}
\label{e4int}
\left( \frac{d~}{d\tau}\right)^3 \frac{1}{(2 \pi i)^3} {\rm \mathcal{L}i}_3(1,q)= \frac{1}{120}  E_4(\tau) .
\end{gather}
and
\begin{gather*}
\left( \frac{\partial~}{\partial z}\right)^2 f(z,\tau)  =  -\frac{1}{2 \pi i}  \log\left\{ \frac{\vartheta_1(z,\tau)}{\eta(\tau)}\right\} .
\end{gather*}
Thus the elliptic-trilogarithm may be thought of as a classical function (or, at least, a
neoclassical function) as it may be obtained from classical elliptic functions via nested integration
and other standard procedures. It does, however, provide a systematic way to deal with the
arbitrary functions that would appear this way.

The following proposition describes the fundamental transformation properties of the function: these will be used
in subsequent section to prove the various dif\/ferential and functional identities. The precise def\/initions and normalizations
of the various objects used here are given in Appendix~A. Also, the notation $F \simeq G$ will be used if the functions
$F$ and $G$ dif\/fer by a~quadratic function in the variables $z$ and $\tau$. These quadratic terms may be easily
derived, but will play no part in the rest of the paper.

\begin{proposition}\label{trans} The function $f$ has the following transformation properties:
\begin{gather*}
f(z+1,\tau)  \simeq   f(z,\tau) ;\\
f(z,\tau+1)   =   f(z,\tau) ;\\
f(z+\tau,\tau)   \simeq   f(z,\tau) +  \left\{ \frac{1}{6} z^3 + \frac{1}{4} z^2 \tau + \frac{1}{6} z \tau^2 + \frac{1}{24} \tau^3\right\} ;\\
f(-z,\tau)  \simeq   f(z,\tau) .
\end{gather*}
The function also has the alternative expansions:
\begin{gather}
\label{fseries1}
f(z,\tau)   \simeq    {-\frac{1}{(2 \pi i)} \left\{ \frac{1}{2} z^2 \log z + z^2 \log \eta(\tau) \right\}}
 {+\frac{1}{(2\pi i)^3} \sum_{n=1}^\infty \frac{ (-1)^n E_{2n}(\tau) B_{2n} }{(2n+2)! (2n)}(2 \pi z)^{2n+2}}
\end{gather}
and
\begin{gather}
\label{fseries2}
f(z,\tau)  =  {\frac{1}{(2\pi i)^3} {\rm Li}_3\left(e^{2 \pi i z}\right)+\frac{1}{12} z^3 - \frac{1}{24} z^2 \tau }
 {-\frac{4}{(2\pi i)^3} \sum_{r=1}^\infty \left\{
\frac{q^r}{(1-q^r)} \right\} \frac{\sin^2 (\pi r z)}{r^3}}.
\end{gather}
The first of these imply the following the transformation property:
\[
f\left(\frac{z}{\tau}, -\frac{1}{\tau}\right) \simeq \frac{1}{\tau^2} f(z,\tau) - \frac{1}{\tau^3} \frac{z^4}{4!}.
\]
\end{proposition}

\begin{proof}
The f\/irst three relation follow immediately from the def\/inition. The fourth used the
inversion formula for polylogarithms (\ref{polyloginversion}).

The proof of (\ref{fseries1}) and (\ref{fseries2}) just involves some careful resumming. Consider the f\/irst two terms in the def\/inition
of $f$:
\begin{gather*}
\sum_{n=0}^\infty {\rm Li}_3\big(q^n e^{2 \pi i z}\big) + \sum_{n=1}^\infty {\rm Li}_3\big(q^n e^{-2 \pi i z}\big)    =   {\rm Li}_3\big(e^{2 \pi i z}\big)+  2 \sum_{s=0}^\infty
\frac{(-1)^s}{(2s)!} \left\{ \sum_{n,r=1}^\infty q^{nr} r^{2s-3} \right\} (2 \pi z)^{2s} .
\end{gather*}
From this series (\ref{fseries2}) follows immediately. To obtain (\ref{fseries1}) one rearranges the terms.
The $s=0$ term cancels in the f\/inal expression and the remaining terms may be re-expressed in terms of
Eisenstein series (for $s>1$) or the Dedekind function (for $s=1$). Finally, using the result
\begin{gather*}
\frac{1}{(2 \pi i)^3}\frac{d^3~}{dz^3} {\rm Li}_3\left(e^{2 \pi i z}\right)   =   -\frac{1}{2} \left[1+\coth(\pi i z)\right] =  -\left[ \frac{1}{2} +\frac{1}{(2 \pi z)} +  \sum_{n=1}^\infty \frac{B_{2n}}{(2k)!} (2 \pi i z)^{2k-1}\right]
\end{gather*}
one may obtain a series for ${\rm Li}_3\left(e^{2 \pi i z}\right).$ Putting all these parts together
gives the series~(\ref{fseries1}).
\end{proof}

\section[Differential identities]{Dif\/ferential identities}\label{section3}

\begin{theorem}\label{little} The function $f$ satisfies the equation
\[
f^{(3,0)}   f^{(1,2)} - \big( f^{(2,1)} \big)^2 + \frac{1}{3} f^{(0,3)}=-\frac{1}{144} E_4(\tau) .
\]
\end{theorem}

\begin{proof}
We denote the left hand side of the dif\/ferential equation by $\Delta(f)(z,\tau) $ and study its transformation
properties. It follows from Proposition \ref{trans} that the third derivatives of $f$ are invariant under the transformation $z\mapsto z + 1  $ and that under the transformation $z\mapsto z+  \tau$ one has:
\begin{gather*}
f^{(3,0)}(z+ \tau)   =   f^{(3,0)}(z) + 1  , \\
f^{(2,1)}(z+ \tau)   =   f^{(2,1)}(z) - f^{(3,0)}(z) - \frac{1}{2},\\
f^{(1,2)}(z+ \tau)   =   f^{(1,2)}(z) -2 f^{(2,1)}(z) + f^{(3,0)}(z) + \frac{1}{3} ,\\
f^{(0,3)}(z+ \tau)   =   f^{(0,3)}(z) -3  f^{(1,2)}(z) +3 f^{(2,1)}(z) -  f^{(3,0)}(z) - \frac{1}{4} .
\end{gather*}

It immediately follows that combination $\Delta(f)$ is a doubly periodic function (even though individual
components are not). From the Laurent expansions of~$f$, the only term with a pole is~$f^{(3,0)}$, which
has a simple pole at $z=0$, but this cancels with the zero at~$z=0$ of the term~$f^{(1,2)}$. Thus
$\Delta(f)$ is a doubly period function with no poles, and hence must be independent of~$z$, i.e.\
a $\vartheta$-constant.

Using the series expansion in Proposition~\ref{trans} one f\/inds that
\[
\Delta = \frac{1}{12} \left\{ \frac{1}{2 \pi i} E_2^\prime - \frac{1}{12} E_2^2 \right\}
\]
and hence the result follows on using the Ramanujan identity
\[
q\frac{dE_2}{dq} = \frac{E_2^2-E_4}{12} .
\]
An alternative proof is to study the modularity properties of $\Delta$ under $\tau\rightarrow -\tau^{-1} $.
One f\/inds that $\Delta$ must be a modular form of degree 4, and hence must be a multiple of $E_4 $, the
constant of proportionality being straightforward to calculated. Note, one could easily redef\/ine $f$, using~(\ref{e4int}), so that $\Delta=0 $.
\end{proof}

This dif\/ferential identity contains much information. Using the $z$-expansion of the
function~$f$ it is equivalent to an inf\/inite family of identities between the
Eisenstein series and their derivatives, though these will all be consequences of
the Ramanujan relations and the fact that the ring of modular forms is f\/initely
generated by~$E_4$ and~$E_6$. Alternatively, using the $q$-expansion of~$f$ it
is equivalent to an inf\/inite family of number-theoretic trigonometric identities
(number theoretic since in the $q$-series for~$f$ the~$q^n$ coef\/f\/icient will involve a sum over~$r|n$).

\section{Functional identities and rank 2-root systems}\label{section4} 

The basic identity ({\ref{thetaidentity}) may be associated to the $A_2$-root system
via the identif\/ication $a=(\alpha,{\bf z})$, $b=(\beta,{\bf z})$, $c=-(\alpha+\beta,{\bf z})$,
where $\alpha$ and $\beta$ are the positive roots and $(\cdot,\cdot)$ is the standard Euclidean inner product.
This immediately suggest that there
should be variants of this identity for other rank-2 root systems.

\begin{lemma} Let $\mathcal{R}$ be the root system for the $2$-dimensional Coxeter groups
$A_2$, $B_2$ or $G_2$, with the standard normalization
for $\alpha$, $\beta$ positive simple roots:
\begin{gather*}
A_2 : \quad (\alpha,\alpha)=(\beta,\beta)=2 ,\qquad (\alpha,\beta)=-1 ,\\
B_2 : \quad (\alpha,\alpha)=2 ,\qquad (\beta,\beta)=1 ,\qquad (\alpha,\beta)=-1 ,\\
G_2 : \quad (\alpha,\alpha)=6 ,\qquad (\beta,\beta)=2 ,\qquad (\alpha,\beta)=-3 .
\end{gather*}
Then
\[
\sum_{\alpha\neq\beta\in\mathcal{R}^{+}} (\alpha,\beta) f^{(3,0)}\big(({\bf z},\alpha),\tau\big) \cdot f^{(3,0)}\big(( {\bf z},\beta),\tau\big)
+ \sum_{\alpha\in\mathcal{R}^{+}} k_{\alpha} f^{(2,1)}\big(({\bf z},\alpha),\tau\big) = 0,
\]
where:
\begin{gather*}
\bullet \  \ A_2: \quad \mbox{$k_\alpha =1$ for all roots};\\
\bullet \ \ B_2: \quad k_{\rm short}=2 ,\qquad k_{\rm long}=1;\\
\bullet \ \ G_2:  \quad k_{\rm short}=10 ,\quad k_{\rm long}=6.
\end{gather*}
\end{lemma}
The proof is entirely standard and is omitted: it follows the proof of Theorem \ref{little}; one f\/irst
shows that the left hand side is doubly periodic in all $a$, $b$, $c$ variables (constrained via $a+b+c=0$) with no poles in all
these variables, and hence must be independent of all three variables, i.e.\ it must be a $\vartheta$-constant. Modular
properties then f\/ix this $\vartheta$-constant.

Many other functional identities may be derived using the same ideas. We present here
a~(not exhaustive) list of identities, all of the general form (\ref{genform}), for
the root system $A_2$ and $B_2$ (see Appendix~B)\footnote{The corresponding identities for $G_2$ are
available on request.}.
The
origin of such identities stems from the analysis of elliptic solutions to the
Witten--Dijkgraaf--Verlinde--Verlinde equation. This application will be presented
in the next section.

\subsection[$A_2$-identities]{$\boldsymbol{A_2}$-identities}\label{section4.1}

\begin{gather}
\left\{
\begin{array}{c}
\displaystyle{f^{(3,0)}(x + y) \big[f^{(2,1)}(x) - f^{(2,1)}(y)\big]}\vspace{2mm}\\
+\displaystyle{f^{(3,0)}(y)\big[ f^{(2,1)}(x + y) - f^{(2,1)}(x)\big]}
\end{array}
\right\}\nonumber\\
\qquad{}+\left\{f^{(1,2)}(x) - \frac{1}{2} f^{(1,2)}(y) + \frac{1}{2} f^{(1,2)}(x + y)\right\}=0,
\label{A2b}
\\
\left\{
\begin{array}{c}
 \displaystyle{f^{(3,0)}(x)\big[f^{(1,2)}(x + y) - f^{(1,2)}(y)\big]}\vspace{2mm}\\
+\displaystyle{f^{(3,0)}(y)\big[f^{(1,2)}(x + y) - f^{(1,2)}(x)\big]}\vspace{2mm}\\
- \tfrac{2}{3}
\displaystyle{f^{(3,0)}(x + y)\big[f^{(1,2)}(x) + f^{(1,2)}(y)\big]}
\end{array}
\right\}
 +
\left\{
\begin{array}{c}
 \frac{2}{3} f^{(2,1)}(x + y)f^{(2,1)}(x)\vspace{2mm}\\
+ \tfrac{2}{3} f^{(2,1)}(x + y)f^{(2,1)}(y) \vspace{2mm}\\
 - \tfrac{8}{3} f^{(2,1)}(x) f^{(2,1)}(y)
\end{array}
\right\} \nonumber\\
\qquad{}+ \frac{10}{9}  f^{(0,3)}(x + y)  +   \frac{1}{108}  E_4(\tau)=0 .\nonumber
\end{gather}
The dependence of $f$ on $\tau$ in these formulae has been suppressed for
notational convenience.

\section{Applications}\label{section5}

In this section we present two applications of these identities: one is in the theory of
elliptic solutions to the Witten--Dijkgraaf--Verlinde--Verlinde (or WDVV) equations of topological
quantum f\/ield theory \cite{strachan}, and the second, more speculative and incomplete, in the theory
of elliptic Dunkl-type operators.

\subsection{Frobenius manifolds and solutions to the WDVV-equations}\label{section5.1}

The basic identity (\ref{thetaidentity}) has particularly simple rational and
trigonometric limits.
Given $a ,b ,c \in\mathbb{C}$ such that $a+b+c=0$ then the rational limit
gives
\begin{gather}
\frac{1}{a} \cdot \frac{1}{b} +\frac{1}{b} \cdot \frac{1}{c} +\frac{1}{c} \cdot \frac{1}{a} =0
\label{rationallimit}
\end{gather}
and the trigonometric limit gives
\begin{gather}
\cot(a) \cot(b) +\cot(b) \cot(c) +\cot(c) \cot(a) =1 .
\label{triglimit}
\end{gather}
These identities are well known, and one application of them is in the construction of
solutions to the WDVV-equation. In particular, the
function
\[
F=\frac{1}{4} \sum_{\alpha \in \mathcal{R}_W} (\alpha,{\bf z})^2 \log (\alpha,{\bf z})^2
\]
satisf\/ies the WDVV equation (see, for example, \cite{ChalVes,Marsh})
\[
[\mathbf{F}_i,\mathbf{F}_j]=\mathbf{0},
\]
where
\[
\left(\mathbf{F}_i\right)_{j}^{k}=g^{kr}\frac{\partial^3F}{\partial z^i\partial z^j \partial z^r} ,
\]
(here $g$ is the Euclidean metric) and $\mathcal{R}_W$ are the roots of a Coxeter group $W $. The verif\/ication of this
result reduces to sums over vectors $\mathcal{R}_W \cap \mathfrak{U}$, where $\mathfrak{U}$ is
an arbitrary 2-plane, and the only non-trivial conf\/igurations are the 3 irreducible 2-dimensional root systems.
The WDVV equations are then satisf\/ied by virtue of identities such as (\ref{rationallimit}).
Similarly, the verif\/ication that the trigonometric function~\cite{misha,MH}
\[
F={\rm cubic~terms} + \sum_{\alpha \in \mathcal{R}_W} h_{\alpha} {\rm Li}_3\big(e^{i (\alpha,{\bf x})}\big)
\]
satisf\/ies the WDVV equation involves use of identities of the form (\ref{triglimit}). It is not
therefore surprising that the elliptic identities of the form (\ref{thetaidentity}) should play a role in elliptic solutions
of the WDVV equation. Indeed, the origin of the dif\/ferential equation in Theorem \ref{little} is the following:

\begin{lemma}\label{A1solutionA} The function
\begin{gather*}
F(u,z,\tau)   =   \frac{1}{2} u^2 \tau - u z^2 + h(z,\tau)
\end{gather*}
satisfies the WDVV equations if and only if $h(z,\tau)$ satisfies the
partial differential equation
\[
h^{(3,0)}  h^{(1,2)}  - \big(h^{(2,1)}  \big)^2 + \frac{1}{3}  h^{(0,3)}  =0 .
\]
\end{lemma}

From Theorem \ref{little}, one solution of this equation is clearly
\begin{gather*}
{h}(z,\tau)=f(z,\tau) + \frac{5}{2(2\pi i)^3} {\rm \mathcal{L}i}_3(1,q) .
\end{gather*}
However, another solution is
\[
h(z,\tau)= f(2z,\tau)- 4 f(z,\tau)
\]
and it is this solution that corresponds to a nice geometric structure, namely  the dual
prepotential to the $A_1$-Jacobi group orbit space \cite{RS}. This, and the
corresponding potentials that can be calculated for the $A_N$-Jacobi group orbit \cite{RS} space suggest
that one can develop the following functional ansatz
for more general class of solutions:
\begin{gather}
F(u,{\bf z}, \tau) = \frac{1}{2} u^2 \tau - \frac{1}{2} u ({\bf z},{\bf z}) +  \sum_{\alpha\in\mathfrak{U}} h_{\alpha} f\big(({\bf z},\alpha), \tau\big) .
\label{defF}
\end{gather}
Here $(u,{\bf z},\tau) \in \mathbb{C}\times\mathbb{C}^N\times\mathbb{H}$ and $\mathfrak{U}$ is a certain f\/inite set of \lq root vectors\rq~in
$\mathbb{C}^N$. Note that the vector f\/ield $e=\frac{\partial~}{\partial u}$ is the unity
of the multiplication, and that
\[
g=2 du\, d\tau - (d{\bf z},d{\bf z}) ,
\]
where $(~,~)$ is the standard complexif\/ied Euclidean inner product on $\mathbb{C}^N $.
The precise conditions on the
set of vectors $\mathfrak{U}$ which are required for (\ref{defF}) to satisfy the WDVV equations were derived in~\cite{strachan}. These start
with the def\/inition of a complex Euclidean $\vee$-system:

\begin{definition}[\cite{FV2}]
Let $\mathfrak{h}$ be a complex vector space with non-degenerate
bilinear form $(\,,\,)$ and let $\mathfrak{U}$ be a collection of
vectors in $\mathfrak{h}$. A complex Euclidean $\vee$-system
$\mathfrak{U}$ satisf\/ies the following conditions:
\begin{itemize}\itemsep=0pt
\item{} $\mathfrak{U}$ is well distributed, i.e.\
$\sum\limits_{\alpha\in\mathfrak{U}} h_\alpha (\alpha,{\bf u})
(\alpha,{\bf v}) = 2 h^\vee_\mathfrak{U} ({\bf u},{\bf v})$ for
some $\lambda$;

\item{} on any 2-dimensional plane $\Pi$
the set $\Pi\cap\mathfrak{U}$ is either well distributed or
reducible (i.e.\ the union of two non-empty orthogonal subsystems).
\end{itemize}
\end{definition}

  With this one can def\/ine an elliptic $\vee$-system as follows:

\begin{definition}\label{ellipticVee} An elliptic $\vee$-system $\mathfrak{U}$ is a complex Euclidean $\vee$-system with
the following additional conditions:
\begin{itemize}\itemsep=0pt
\item{} $\sum\limits_{\alpha \in\mathfrak{U}} h_{\alpha} (\alpha,{\bf z})^4 = 3({\bf z},{\bf z})^2$;
\item{} the three conditions:
\begin{gather*}
\sum_{\beta \in \Pi_\alpha \cap \mathfrak{U}} h_\beta (\alpha,\beta)(\beta,\alpha^\perp)^n=0 ,\\
\sum_{\beta \in \Pi_\alpha \cap \mathfrak{U}} h_\beta (\alpha,\beta) (\alpha \wedge \beta) (\beta,\alpha^\perp)^n=0 ,\\
\sum_{\beta \in \Pi_\alpha \cap \mathfrak{U}} h_\beta (\alpha,\beta) (\alpha \wedge \beta)\otimes (\alpha \wedge \beta)(\beta,\alpha^\perp)^n=0 ;
\end{gather*}
\item{} there exists a full $N$-dimensional weight lattice of vectors ${\bf p}$ such that
$({\bf p},\alpha) \in\mathbb{Z}$ for all $\alpha\in \mathfrak{U}\,$.
\end{itemize}
\end{definition}

With this
def\/inition of an elliptic $\vee$-system one can state the main theorem of \cite{strachan}.

\begin{theorem}\label{BigTheorem} Let $\mathfrak{U}$ be an elliptic $\vee$-system. If $h^\vee_\mathfrak{U}=0$ then the function
\begin{gather*}
F(u,{\bf z}, \tau) = \frac{1}{2} u^2 \tau - \frac{1}{2} u ({\bf z},{\bf z}) +  \sum_{\alpha\in\mathfrak{U}} h_{\alpha} f\left(({\bf z},\alpha), \tau\right)
\end{gather*}
satisfies the WDVV equations. If $h^\vee_\mathfrak{U}\neq 0$ then the modified prepotential
\begin{gather*}
F\longrightarrow F+ \frac{10 \left( h^\vee_\mathfrak{U}\right)^2 }{3(2\pi i)^3} {\rm \mathcal{L}i}_3(1,q)
\end{gather*}
satisfies the WDVV equations.
\end{theorem}

The proof of this theorem is entirely analogous to the proof of Theorem~\ref{little}: the f\/irst and third
parts of Def\/inition \ref{ellipticVee} ensure that that functions $\mathbf{\Delta}=[\mathbf{F}_i,\mathbf{F}_j]$,
the obstructions to the WDVV-equations from holding,
are doubly periodic in all ${\bf z}$-variables, and the second conditions ensure that these
functions have no poles and hence must be a $\vartheta$-constant. Modular properties then f\/ix the $\vartheta$-constant as above, and hence
to the vanishing of the obstructions $\mathbf{\Delta}$.

One unexpected result is the a \lq pure\rq~root system does not suf\/f\/ice: it is not, in general,
an elliptic $\vee$-system. Thus generalization of classical integrable systems from rational
and trigonometric to elliptic based purely on Coxeter root systems may not always work.

\subsection{Dunkl-type operators}\label{section5.2}

There is a close similarity between the mathematics behind the WDVV equations and the commutativity
of Dunkl operators. Given the role of the above functional identities in the construction of
solutions to the associativity equations a natural question to raise is whether or not they have a role
in Dunkl-type operators. The results presented here will be for the Coxeter group $A_2$ alone.

Let
\begin{gather*}
\Xi^{(-1)}(\xi)   =   \partial_\xi + \sum_{a \in \mathfrak{U}} k_a (a,\xi) {\hat{s}}_a ,\\
\Xi^{(i)}(\xi)   =   \partial_\xi + \sum_{a \in \mathfrak{U}} k_a (a,\xi) f^{(3-i,i)}({\bf z},a) {\hat{s}}_a ,\qquad i=0,1,2,3 .
\end{gather*}
Here $\mathfrak{U}=\{\alpha,\beta,-(\alpha+\beta)\}$ where $\alpha$ and $\beta$ are the simple positive
roots of $A_2$. The following follows from direct calculation, on using the basic identity (\ref{thetaidentity}).
\begin{proposition}
Let
\[
\mathbb{F}^{(a,b)}(\xi,\eta)=
\big[ \Xi^{(a)}(\xi),\Xi^{(b)}(\eta) \big] + \big[ \Xi^{(b)}(\xi),\Xi^{(a)}(\eta) \big] .
\]
Then
\begin{gather*}
\mathbb{F}^{(0,-1)}(\xi,\eta)   =   0,\\
\mathbb{F}^{(0,0)}(\xi,\eta) + \mathbb{F}^{(1,-1)}(\xi,\eta) = 0 .
\end{gather*}
\end{proposition}

(Note that the second of these can be written more succinctly as
\[
g_{\mu\nu} \big[ \Xi^{(\mu)}(\xi), \Xi^{(\nu)}(\eta)\big]=0
\]
where $g=du d\tau + 2 dz^2$.) These relations bears a close similarity with the zero-curvature
relations of the elliptic WDVV equations. However, terms such as $\mathbb{F}^{(0,1)}(\xi,\eta)$ cannot
be expressed in terms of other $\mathbb{F}^{(i,j)}$, which might have been expected via the use of
the generalized Frobenius--Stickelberger relation~(\ref{A2b}).

These operators $\Xi$ do have simple rational and trigonometric limits, from which one can
recover standard results. In the rational limit $f^{(3,0)}(z,\tau) \sim z^{-1}$ and one hence obtains the original result \cite{Dunkl} of Dunkl
\[
 \big[ \Xi^{(0)}(\xi),\Xi^{(0)}(\eta) \big] = 0 ,
\]
and in the trigonometric limit ($q\rightarrow 0$) $f^{(2,1)}(z,\tau) \rightarrow -\frac{1}{12}$ and so
\[
\Xi^{(1)}(\xi)\rightarrow -\frac{1}{12} \Xi^{(-1)}(\xi) .
\]
With this one obtains Heckman's operators~\cite{Heckman} which, following Cherednik, can be made into
a pair of commuting operators~\cite{Cherednik}.

It does appear, based on the above calculations for $A_2$, that these identities cannot be used
to construct new classes of elliptic-type Dunkl operators (thought the basic identity (\ref{thetaidentity}) does
play a role in elliptic KZ-theory~\cite{Etingof}). This is, however, scope for further investigation of this
problem.

\section{Conclusions}\label{section6}

One recurrent theme in the theory of integrable systems is the tower of
generalizations
\[
{\rm rational} \ \longrightarrow \ {\rm trigonometric} \ \longrightarrow \ {\rm elliptic},
\]
and we have seen in this paper how the Frobenius--Stickelberger relation, via its $\vartheta$-function
form, ref\/lects this. The WDVV-equation are themselves an example of an integrable system; they
arise as the zero-curvature conditions of the connection \cite{dubrovin1}
\[
{\tilde\nabla}_X Y = \nabla_X Y + \lambda X \circ Y ,
\]
and the solutions described in Theorem (\ref{BigTheorem}) (at least for the Weyl groups $A_N$ and $B_N$, and
conjecturally for all Weyl groups) sit at the right of
the following tower of generalizations:
\[
\underset{\left\{
\begin{array}{c}{\rm Coxeter~group}\\{\rm orbit~space} \end{array}
\right\}}{\mathbb{C}^N/W}  \  \longrightarrow  \
\underset{\left\{
\begin{array}{c}{\rm Extended~af\/f\/ine~Weyl}\\{\rm orbit~space} \end{array}
\right\}}{\mathbb{C}^{N+1}/{\widetilde{W}}} \  \longrightarrow  \
\underset{\left\{
\begin{array}{c}{\rm Jacobi~group}\\{\rm orbit~space} \end{array}
\right\}}{\Omega/J(\mathfrak{g})} .
\]
\noindent (technically the solutions are the corresponding \lq almost dual\rq~solutions \cite{dubrovin2} to the Frobenius
manifold structures on these spaces and the solutions described in Theorem (\ref{BigTheorem}) correspond to the
Frobenius manifolds constructed in \cite{B, Satake}). These Jacobi group orbit spaces use in their construction properties of Jacobi forms
\cite{EZ,wirth}, functions which may be thought of as elliptic generalizations of $W$-invariant polynomials.

This tower of generalizations clearly need not stop at elliptic solutions. The $A_N$-Frobenius manifolds for the
rational (indeed polynomial), trigonometric and elliptic cases have a Hurwitz space description in terms of the moduli
space $H_{g,N}(k_1,\ldots, k_l)$ of branched coverings of the Riemann sphere. An interesting
question is whether or not there is an orbit space construction for these more general spaces:
\[
\begin{array}{ccccccccc}
H_{0,N}(N) & \longrightarrow & H_{0,N}(k,N-k) & \longrightarrow & H_{1,N}(N) & \longrightarrow & \cdots & \longrightarrow &H_{g,N}(N) \vspace{3mm}\\
\updownarrow&  & \updownarrow &  &  \updownarrow & & & & \updownarrow\vspace{3mm}\\
\mathbb{C}^N/A_N & \longrightarrow & \mathbb{C}^{N+1}/{\widetilde{A}_N^{(k)}} & \longrightarrow & \Omega/J({A_N})& \longrightarrow & \cdots & \longrightarrow &
\begin{array}{c}
{\rm orbit}\\
{\rm space}\\
{\rm structure?}
\end{array}
 \end{array}
\]
It seems sensible to conjecture that such an orbit space exists. One would expect Seigel modular forms to play a role
instead of the modular forms used here. Higher genus Jacobi forms certain have been studied, but their use has yet to
percolate into the theory of integrable systems. The development, and applications of, the neo-classical $\vartheta$-function identities studied in Section~\ref{section4} remains to be done systematically. Higher genus analogues of these identities certain exist,
since there exists almost-dual prepotentials on these Hurwitz spaces which, by construction, satisfy the WDVV equations.
In the genus~0 and genus~1 cases, the prepotential is very closely related to the prime form on the Riemann surface.
This may be the starting point for the development of a functional ansatz for the higher genus cases. Central to the
results presented here are the quasi-periodicity and modularity properties of the elliptic polylogarithm, and these
were obtained from the analytic properties of this function; the only role the analytic properties play were in the
development of these transformation properties. It would be attractive if one could obtain these directly from the
geometric properties of the prime form. This approach could then be used in the higher genus case where the analytic
properties are likely to be considerably more complicated.

\pdfbookmark[1]{Appendix A}{appendixA}

\section*{Appendix A}}

There are, unfortunately, many dif\/ferent def\/initions and
normalizations for elliptic, number-theoretic and other special functions. Here
we list the def\/initions used in this paper. Let $q=e^{2\pi i \tau}$, where $\tau \in\mathbb{H}$.

\begin{itemize}\itemsep=0pt

\item{} $\vartheta_1$-function:
\[
\vartheta_1(z|\tau) = -i \left(e^{\pi i z}-e^{-\pi i z}\right) q^{\frac{1}{8}} \prod_{n=1}^\infty
(1-q^n) \left(1-q^n e^{2 \pi i z}\right)\left(1-q^n e^{-2\pi i z}\right) .
\]
The fundamental lattice is generated by $z\mapsto z+1$, $z\mapsto z+\tau$ and the function
satisf\/ies the complex heat equation $\vartheta_1^{\prime\prime}=4 \pi i \vartheta_{1,\tau}$.

\item{} Bernoulli numbers:
\[
\frac{x}{e^x-1} = \sum_{n=0}^\infty B_k \frac{x^n}{n!}.
\]

\item{}Eisenstein series:
\[
E_k(\tau) = 1 - \frac{2k}{B_k} \sum_{n=1}^\infty \sigma_{k-1}(n) q^n,\qquad k \in 2\mathbb{N},
\]
where $\sigma_k(n)=\sum_{d|n} d^k$.

\item{}Dedekind $\eta$-function:
\[
\eta(\tau) = q^{\frac{1}{24}} \prod_{n=1}^\infty \big(1-q^n\big).
\]

\item{} Polylogarithm function:
\[
{\rm Li}_N(z) = \sum_{n=1}^\infty \frac{z^n}{n^N} , \qquad |z|<1.
\]
\end{itemize}

Note that $\vartheta_1$, $E_2$ and $\eta$ are related:
\[
\frac{ \eta^\prime(\tau)}{\eta(\tau)} = \frac{2 \pi i}{24}  E_2(\tau) =
\frac{1}{12 \pi i} \frac{\vartheta_1^{\prime\prime\prime}(0,\tau)}{\vartheta_1^\prime(0,\tau)}.
\]
These have the following properties under inversion of the independent variable:
\begin{gather*}
\tau^{-n} E_n\left(-\frac{1}{\tau}\right)   =  E_n(\tau)  , \qquad n\geq 4 ; \\
\tau^{-2} E_2\left(-\frac{1}{\tau}\right)   =   E_2(\tau) + \frac{12}{2 \pi i \tau} ;\\
\eta\left(-\frac{1}{\tau}\right)   =   \sqrt{\frac{\tau}{i}}  \eta(\tau),
\end{gather*}
where in the last formula the square-root is taken to have non-negative real part. The
poly\-logarithm has the inversion property:
\begin{gather}
(-1)^{N-1} {\rm Li}_N(z^{-1}) = {\rm Li}_N(z) + \sum_{j=0}^N \frac{B_j (2\pi \sqrt{-1})^j}{(N-j)!j!} (\log z)^{N-j} .
\label{polyloginversion}
\end{gather}
This may be used to analytically continue the function outside the unit disc to a multi-valued
holomorphic function on $\mathbb{C}\backslash \{0,1\} $. For a discussion of the monodromy
of the polylogarithm function see~\cite{Ram}.

\pdfbookmark[1]{Appendix B}{appendixB}
\section*{Appendix B}

Here we present the identities corresponding to the $B_2$-root system.
There are more than one identity in each of these two sets: this stems from the facts that
the root systems have roots of dif\/ferent lengths.

\noindent{\bf Set B$\boldsymbol{{}_2}$(a):}
\begin{gather*}
\left\{
\begin{array}{c}
\displaystyle{f^{(3,0)}(x) \big[f^{(2,1)}(y) - f^{(2,1)}(x+y)\big]}\vspace{2mm}\\
{}+\displaystyle{f^{(3,0)}(x+y)\big[ f^{(2,1)}(x) - f^{(2,1)}(x+2y)\big]}\vspace{2mm}\\
{}+\displaystyle{f^{(3,0)}(x+2y)\big[ f^{(2,1)}(x+y) - f^{(2,1)}(y)\big]}
\end{array}
\right\}\\
\qquad{} +\left\{\frac{1}{2}f^{(1,2)}(x) - \frac{1}{2} f^{(1,2)}(x+2y) -2 f^{(1,2)}(y)\right\}=0,
\\
\left\{
\begin{array}{c}
\displaystyle{f^{(3,0)}(y) \big[f^{(2,1)}(x+2y) - f^{(2,1)}(x)\big]}\vspace{2mm}\\
{}+\displaystyle{f^{(3,0)}(x+y)\big[ f^{(2,1)}(x) - f^{(2,1)}(x+2y)\big]}\vspace{2mm}\\
{}+\displaystyle{2 f^{(3,0)}(x+2y)\big[ f^{(2,1)}(x+y) - f^{(2,1)}(y)\big]}
\end{array}
\right\}\\
\qquad{}+\big\{f^{(1,2)}(x) +2 f^{(1,2)}(x+y) -2 f^{(1,2)}(y)\big\}=0.
\end{gather*}

\noindent{\bf Set B$\boldsymbol{{}_2}$(b):}
\begin{gather*}
\left\{
\begin{array}{c}
\phantom{+}
f^{(3,0)}(x) \displaystyle{\big[f^{(1,2)}(x + y) - f^{(1,2)}(y)\big]}\vspace{2mm} \\
{}+f^{(3,0)}(x + y) \displaystyle{\big[f^{(1,2)}(x) + f^{(1,2)}(x + 2y)\big]}\vspace{2mm} \\
{}+f^{(3,0)}(x + 2y) \displaystyle{\big[f^{(1,2)}(x + y) + f^{(1,2)}(y)\big]}
\end{array}\right\} -
2
\left\{
\begin{array}{c}
\displaystyle{f^{(2,1)}(x + y) f^{(2,1)}(x)}\vspace{2mm} \\
{}+\displaystyle{f^{(2,1)}(x + y) f^{(2,1)}(x + 2y)}
\end{array}
\right\} \\
\qquad{}+\frac{1}{3} \left\{
\begin{array}{c}
f^{(0,3)}(x) + 2f^{(0,3)}(x + y) \vspace{2mm} \\
{}+  f^{(0,3)}(x + 2y) + 6f^{(0,3)}(y)
\end{array}
\right\} +  \frac{1}{36} E_4(q)=0 ,\\
\left\{
\begin{array}{c}
\displaystyle{f^{(3,0)}(y)\big[f^{(1,2)}(x + 2y) - f^{(1,2)}(x)\big]}\vspace{2mm} \\
{}+\displaystyle{4 f^{(3,0)}(x+2y)\big[f^{(1,2)}(y) + f^{(1,2)}(x+y)\big]}\vspace{2mm} \\
{}+ \displaystyle{f^{(3,0)}(x + y)\big[f^{(1,2)}(x) + f^{(1,2)}(x+2y)\big]}
\end{array}
\right\}  -  4\left\{
\begin{array}{c}
  \displaystyle{ f^{(2,1)}(x + 2 y)f^{(2,1)}(x+y)}\vspace{2mm} \\
{}+\displaystyle{ f^{(2,1)}(x + 2 y)f^{(2,1)}(y)}
\end{array}
\right\}\\
\qquad{} +\frac{1}{3} \left\{
\begin{array}{c}
\phantom{+}
f^{(0,3)}(x)+ 8 f^{(0,3)}(x + y)\vspace{2mm} \\
+ f^{(0,3)}(x + 2y)+ 8 f^{(0,3)}(y)
\end{array}
\right\}  + \frac{1}{18} E_4(\tau)=0 .
\end{gather*}

\subsection*{Acknowledgments}

I would like to thank Harry Braden, who f\/irst showed me that~(\ref{thetaidentity}) was just the
Frobenius--Stickelberger identity~(\ref{fs}), and Misha Feigin and Andrew Riley for their comments and remarks.

\pdfbookmark[1]{References}{ref}
\LastPageEnding

\end{document}